\documentclass[journal]{IEEEtran}

\usepackage{cite}
\usepackage{amsmath,amssymb,amsfonts}
\usepackage{graphicx}
\usepackage{textcomp}
\usepackage{xcolor}
\usepackage{booktabs}
\usepackage{multirow}
\usepackage{float}
\usepackage{microtype}
\usepackage{hyperref}

\hypersetup{colorlinks=true,linkcolor=black,citecolor=black,urlcolor=black}

\begin{document}

\title{No Convergence in Learned Eavesdropping--Detection
Dynamics for BB84 Quantum Key Distribution}

\author{Ramon Jose C. Bagunu\\
Department of Physical Sciences and Mathematics, University of the Philippines Manila, Ermita, Manila 1000, Philippines\\
\textit{rcbagunu@up.edu.ph}}

\maketitle

\begin{abstract}
In prior work we showed that a reinforcement-learning eavesdropper can intercept BB84 quantum key distribution while holding the quantum bit error rate below the abort threshold and keeping a usable distillable key. In this paper we ask whether an equally adaptive learning detector can close this stealth window. We cast the interaction as an alternating best-response game between a tabular eavesdropper and a tabular threshold-adapting detector. We then measure exploitability, which is the detection rate and distillable key fraction achieved by a fresh attacker best-responding to the final detector. Across three learning rules (Q-Learning, SARSA, Double Q-Learning), three channel-noise levels, and a uniform false-alarm discipline, the dynamics do not settle out to twenty rounds. The exploitability does not approach a fixed point, and its round-to-round swings do not decay. Two controls, one retraining the attacker against a frozen defender and one retraining the defender against a frozen attacker, show that the swings on each side match the variance of the best-response procedure itself. The dynamics fluctuate at the scale of best-response variance rather than converging, with neither player settling to a fixed response. The behavior is noise-dependent, and the stealth window stays open at the low noise where practical QKD operates. The defender contains the attacker on average at intermediate noise without the contest settling, and at high noise the channel rather than the learned policy suppresses the attack. As a candidate mechanism we examine a last-mover information asymmetry, supported by significant anti-persistence at intermediate noise and consistent with best-response noise at low noise. The results suggest that a learned detector cannot be certified from any single round, and that closing the window calls for breaking the information structure rather than out-adapting the attacker.
\end{abstract}


\section{Introduction}
\label{sec:intro}

\IEEEPARstart{Q}{uantum} key distribution (QKD) lets two parties, conventionally Alice and Bob, establish a shared secret key whose security rests on physical laws rather than on assumptions about an adversary's computational power \cite{bennett1984quantum, gisin2002quantum}. In the BB84 protocol \cite{bennett1984quantum}, the no-cloning theorem \cite{wootters1982single} guarantees that any attempt by an eavesdropper, Eve, to measure the quantum signals disturbs them and raises the quantum bit error rate (QBER) that Alice and Bob observe. Security proofs establish a threshold QBER of approximately $11\%$ for BB84, below which a secret key can still be distilled and above which the parties abort \cite{shor2000simple, mayers2001unconditional, lo1999unconditional, tomamichel2012tight}. This threshold is the foundation of practical QKD, since the protocol trusts that an eavesdropper powerful enough to compromise the key necessarily drives the QBER past the abort point \cite{gottesman2004security, scarani2009security}.

A persistent concern is whether this trust is well placed when the eavesdropper is adaptive. Rather than attacking every transmission, a strategic Eve may intercept only a fraction of the signals which trades a smaller amount of stolen information for a QBER that stays below the abort threshold. Our previous work \cite{bagunu2026adversarial} has shown that reinforcement learning (RL) sharpens this threat. There, we modeled Eve as an RL agent that learns, from the publicly broadcast QBER alone, when to attack and when to stay silent. The learned policy holds the induced QBER near the channel-noise floor and keeps a usable distillable key while a fixed-rate attacker operating just below threshold does not. A strategic eavesdropper is therefore a concrete threat, and it is one that the fixed $11\%$ abort threshold could not close the window against. We re-establish the magnitude of that open window independently in the present work, under greedy evaluation of a freshly trained attacker (Sec.~\ref{sec:results-phase}). This result is part of a broader turn toward machine learning in quantum communication, which spans key-rate estimation, protocol optimization, and quantum-enhanced learning agents \cite{nguyen2023deep, behera2025qnnqrl, decker2024qkd, reiss2023deep}.

If a learning attacker can evade a fixed detector, the natural defense is a detector that also adapts. One might expect that an equally capable, learning defender, one that adjusts its decision rule in response to the attacker's behavior, should be able to close the stealth window that the static threshold leaves open. This is the question that we address in this paper. When both the eavesdropper and the detector learn, does the defender win, or does the attacker's advantage persist?

We argue that this question is not well posed as a contest with a winner, because the attacker and an adaptive detector form a coupled learning system rather than a fixed target and a learner chasing it. When the detector adapts to the current attacker, the attacker can re-adapt to the new detector, and so on. This is precisely the setting of best-response dynamics in game theory, where it has long been known that iterated best responses need not converge to a stable equilibrium and can instead fail to settle \cite{brown1951iterative, shapley1964some, fudenberg1991game}. The same non-stationarity is a well-documented obstacle in multi-agent reinforcement learning \cite{busoniu2008comprehensive}. That such non-convergence is possible in the abstract is known, and it is not our claim. Whether this particular interaction fails to settle, how that behavior is organized across channel noise, and what it means for certifying a detector are not implied by the general theory, and these are what we establish. The recognition that the security question the field has posed as a contest with a winner is, for this interaction, better posed as a question about the learning dynamics is itself part of the contribution. Exploitability, meaning how much a fresh best-responding opponent can still gain against a fixed strategy, has become the standard way to measure who, if anyone, holds the advantage \cite{heinrich2016deep, lanctot2017unified}.

In this work we cast the BB84 eavesdropping and detection problem as an alternating best-response game between a learning attacker and a learning, threshold-adapting detector, both realized as tabular RL agents on the BB84 simulator used in prior work. We train the two players against each other over successive rounds and, at each round, measure the exploitability of the current detector against a freshly trained attacker. This is reported in both detection rate and distillable key fraction. Our central finding is that the game does not settle. Out to twenty best-response rounds, the exploitability does not converge. Its round-to-round swings do not decay, and at intermediate noise they are as large in the second half of the run as in the first. Two controls, one that retrains the attacker against a single frozen defender and one that retrains the defender against a single frozen attacker, show that the swings on each side are of the same magnitude as the best-response procedure's own variance. The dynamics fluctuate at the scale of best-response variance rather than approaching a fixed point, with neither player settling to a fixed response. Reactive detection, however adaptive, cannot resolve the contest. It can only trade places with the attacker. As a candidate mechanism we examine a last-mover information asymmetry. Because the QBER is broadcast publicly, whichever player adapts to it last may hold a transient advantage. The trajectory statistics support this reading at intermediate noise, where the exploitability is significantly anti-persistent, and leave it open at low noise, where the fluctuation is indistinguishable from the variance of the best-response procedure itself.

The contributions of this paper are as follows.
\begin{enumerate}
    \item We formulate adaptive eavesdropping and its detection in BB84 as an alternating best-response game between two learning agents, with exploitability (in detection rate and distillable key fraction) as the solution-quality metric.

    \item We show, across three learning rules (Q-Learning, SARSA, and Double Q-Learning), three channel-noise levels, and a uniform false-alarm discipline on the defender, that the game does not settle under best-response dynamics out to twenty rounds, with the exploitability failing to approach a fixed point. We confirm this is genuine non-convergence rather than slow convergence by showing that the swing amplitude does not decay across rounds. Two controls, one per player, show that the residual motion on each side is at the scale of the best-response procedure's own variance. Neither player converges to a fixed response.

    \item We characterize a noise-dependent phase structure. Here the stealth window stays open at low noise, the attacker is contained in the round-averaged outcome at intermediate noise even though round to round the advantage still changes hands rather than settling, and the channel dominates at high noise. We further show that at a fixed noise level the attacker’s learning rule governs whether the dynamics settle or keep fluctuating.

    \item We examine a last-mover information asymmetry intrinsic to the broadcast of the QBER as the candidate cause. This is supported by significant anti-persistence at intermediate noise and left open at low noise. If it is what drives the instability, it is independent of the particular detector lever, which argues that out-adapting the attacker is the wrong defensive goal.
\end{enumerate}

The remainder of this paper is organized as follows. Section~\ref{sec:related} reviews eavesdropping detection in BB84, machine learning for QKD, adaptive eavesdropping, and the game-theoretic background. Section~\ref{sec:methodology} describes the simulator, the learning attacker and defender, the false-alarm discipline, and the alternating best-response game. Section~\ref{sec:results} presents the exploitability results, the phase structure, and the convergence analysis. Section~\ref{sec:discussion} discusses the last-mover asymmetry and its implications, and then we conclude at Section~\ref{sec:conclusion}.

\section{Related Work}
\label{sec:related}

This paper sits at the intersection of four lines of research, namely eavesdropping detection in BB84, machine learning applied to quantum key distribution, the treatment of eavesdroppers as adaptive learning agents, and the study of learning dynamics in games. We review each in turn, with emphasis on the assumptions they make about the adversary. The first two strands assume a fixed or externally specified attacker and ask how well a protocol or detector performs against it. The third relaxes this by letting the attacker learn, but still against a fixed detector. The fourth supplies the best-response dynamics and exploitability, in which the attacker and the detector learn against each other and no fixed target exists on either side.

\subsection{Eavesdropping detection in BB84}
The security of BB84 relies on the quantum bit error rate. Any intercept-resend attack imprints errors that, in principle, betray the eavesdropper once the QBER crosses the distillation threshold \cite{shor2000simple, gisin2002quantum}. Practical detectors operate by comparing the measured QBER against this threshold and aborting when it is exceeded \cite{gottesman2004security, scarani2009security}. These foundations have been extended to realistic devices and to protocol variants designed to remove hardware-level attack surfaces, including photon-number-splitting limits on weak-coherent sources \cite{brassard2000limitations}, measurement-device-independent QKD \cite{lo2012mdi}, and twin-field QKD \cite{lucamarini2018tfqkd}, as well as to network-scale deployment over quantum repeaters \cite{xu2020secure, azuma2023repeaters} and to formal security certification standards \cite{etsi2023qkd016}. Lee \emph{et al.} \cite{lee2022quantum} studied eavesdropping detection in BB84 directly, and established the detection baseline that our line of work builds upon. A recurring assumption in this literature is that the eavesdropper attacks at a fixed rate, so that the induced QBER is a stable signal that the detector can threshold. Our concern is the regime in which that assumption fails, where an eavesdropper varies its attack rate to shape the QBER it presents.

\subsection{Machine learning in quantum key distribution}
Machine learning has been applied across the QKD stack, including reinforcement learning for key distribution over quantum-repeater networks \cite{reiss2023deep}. Reinforcement learning in particular has been used for resource allocation and control in quantum communication \cite{reiss2023deep}, and quantum-enhanced learning agents have been proposed for QKD tasks \cite{behera2025qnnqrl}. More broadly, deep reinforcement learning has become a standard tool in cybersecurity, where the attacker and defender are modeled as learning agents \cite{nguyen2023deep}. These works treat learning as a means of optimizing a protocol or a defense against a fixed adversary, and they do not study what happens when attacker and defender learn against each other.

\subsection{Adaptive and learned eavesdropping}
A smaller but growing body of work considers the eavesdropper itself as an adaptive, learning agent. Our previous work \cite{bagunu2026adversarial} modeled Eve as a tabular RL agent that learns, from the public QBER alone, a partial-interception policy, with a per-qubit reward that pays for correctly measured information and penalizes both basis errors and detection. The learned attacker holds the induced error rate near the channel-noise floor and retains a substantial usable key fraction, which then outperforms any fixed-rate attacker on the stealth-versus-key trade-off. It does so against the static $11\%$ abort threshold, which cannot close the resulting stealth window. The present paper reuses that attacker and that simulator unchanged, and asks what happens when the detector is allowed to learn as well. Closely related, Decker \emph{et al.} \cite{decker2024qkd} cast the optimization of eavesdropping attacks on a quantum-circuit implementation of BB84 as a quantum machine learning task. Tasar \cite{tasar2025adversarial} examines the adversarial limits of quantum certification, which shows that a generative adversary can drive every tested detector to chance-level performance. Nearest to the present setting, Mohamed and Al-Kuwari \cite{mohamed2026adversarial} model intrusion detection in decoy-state QKD as a minimax game between a learning defender and an adaptive adversary, with the defender optimized directly for finite-key secret-fraction retention. They report that the learned defender preserves most of the honest key rate against adaptive attacks. In a related recent line, Mordarski \emph{et al.} \cite{mordarski2026drift} pose adaptive eavesdropping under device drift as a constrained Markov decision process in which the attacker searches the attack circuit jointly over gate structure and angles, and they measure the information an attacker gains by adapting to the drift. Their learner selects a quantum circuit per round rather than modulating a fixed intercept-resend attack against the broadcast QBER as our attacker does, and their question concerns the value of adaptation under drift rather than the convergence of a detection contest. The present work shares the premise that the eavesdropper reading the public record is a sequential learner, but it asks a different question. Rather than reporting a defender that wins a fixed contest, we ask whether the attacker and detector interaction converges at all. We find that for BB84 intercept-resend, it does not. A defender that appears to contain the attacker at one round is contained in turn at the next, so a single favorable snapshot is not evidence of a closed window. We answer this with exploitability measured across best-response rounds.

\subsection{Learning in games and non-convergence}
Coupled learners that may fail to converge is a classical result in game theory. Brown's fictitious play \cite{brown1951iterative} established iterated best-response as a solution procedure, and Shapley \cite{shapley1964some} exhibited games in which it does not converge but cycles indefinitely. This phenomenon is standard in modern treatments \cite{fudenberg1991game}. In multi-agent reinforcement learning, the same coupling makes each agent's environment non-stationary, which is a well-documented obstacle to convergence \cite{busoniu2008comprehensive}. Because a converged solution cannot be assumed, the field measures strategies by their exploitability. This is the gain that a fresh best-responding opponent can still achieve, which underlies self-play and policy-space methods for large games \cite{heinrich2016deep, lanctot2017unified}. We import exploitability into the QKD setting as the natural way to quantify who, if anyone, holds the advantage in the eavesdropping and detection game. We then use its behavior across rounds as direct evidence of non-convergence. That best-response dynamics can fail to converge is classical, and it is not the claim of this paper. Many coupled-learning problems do converge, and whether a given interaction settles or cycles is a property of that specific interaction rather than a generic fact. What is not known in advance is whether the BB84 eavesdropping and detection game converges, and if it does not, how its non-convergence is organized across operating conditions. That is the question we answer, together with the security consequence that a QKD detector cannot be certified by its performance in any single round.

\section{Methodology}
\label{sec:methodology}

We study the attacker and detector interaction on the same qubit-level BB84 simulator used in our prior work, so that the eavesdropper faced here is the same learned adversary whose capability that work established. This section describes the protocol and threat model (Sec.~\ref{sec:methods-bb84}), the learning attacker (Sec.~\ref{sec:methods-atk}), the distillable-key metric (Sec.~\ref{sec:methods-key}), the learning defender (Sec.~\ref{sec:methods-def}), the false-alarm discipline (Sec.~\ref{sec:methods-fpr}), and the alternating best-response game and its exploitability metric (Sec.~\ref{sec:methods-game}).

\subsection{BB84 protocol and threat model}
\label{sec:methods-bb84}
In BB84, Alice encodes each bit in one of two conjugate bases chosen at random, and Bob measures in a randomly chosen basis \cite{bennett1984quantum}. After transmission, the parties publicly compare bases and keep only the matched-basis events (sifting), which retains approximately half the qubits. The channel is modeled with a per-qubit background noise rate $\mu_{ch}$, which represents the baseline error on an unattacked link. The eavesdropper Eve performs an intercept-resend attack, in which on an attacked qubit she measures in a randomly chosen basis and resends her result. When her basis does not match Alice's, which happens half the time, she irreversibly disturbs the state, so that under full interception roughly $25\%$ of Alice and Bob's sifted bits disagree. Alice and Bob accumulate transmissions into blocks of $100$ qubits. At the end of each block, they announce the block QBER over the authenticated public channel, after which a detector decides whether to continue or abort. A standard BB84 detector aborts once the block QBER reaches the distillation threshold of approximately $11\%$, above which privacy amplification can no longer guarantee a secure key \cite{shor2000simple, gisin2002quantum}. The public broadcast of the block QBER is central to this paper since it is the signal on which both players condition, and the order in which they react to it is what creates the asymmetry we analyze in Sec.~\ref{sec:discussion}.

\subsection{The learning attacker}
\label{sec:methods-atk}
We model Eve as a tabular reinforcement-learning agent acting once per qubit, identically to that previous work. Her state encodes three quantities observable to her, namely the QBER announced at the last checkpoint, in $16$ bins of width $1\%$ over $[0,15\%]$, her own attack count in the current block, in $21$ bins of width $5$ over $[0,100]$, and the block index, which gives $16 \times 21 \times 10 = 3360$ reachable states under the fixed $10$-block horizon used throughout. (The simulator allocates $15$ block bins to support longer horizons, and only the first $10$ are reached here.) Her action set is binary, pass ($0$) or attack ($1$). The reward is assigned per qubit,
\begin{equation}
r_t = \begin{cases}
+1  & \text{attack, correct basis (information gained)} \\
-1  & \text{attack, wrong basis (disturbance, no gain)} \\
+2  & \text{checkpoint survived} \\
-50 & \text{detected (QBER} \geq 11\%) \\
0   & \text{pass (no attack)}.
\end{cases}
\label{eq:reward}
\end{equation}
The $+1/{-1}$ information reward makes the immediate expected payoff of a single attack zero, since Eve is right half the time. The agent must therefore learn when to attack rather than merely whether to, while the $+2$ survival bonus rewards multi-block restraint and the dominant $-50$ abort penalty makes evasion the overriding objective. Per-qubit, rather than per-block, reward assignment is essential. This is because aggregating the reward over a block destroys the credit assignment the agent needs to localize profitable interception opportunities.

We consider three value-based learning rules, Q-Learning \cite{watkins1992q}, SARSA \cite{rummery1994line}, and Double Q-Learning \cite{hasselt2010double}. They differ in how to estimate action values and, consequently, in how aggressively they exploit the detector. Q-Learning's off-policy $\max$ operator overestimates action values and yields the most aggressive attacker, Double Q-Learning corrects this bias and is the most conservative, and SARSA's on-policy updates place it between the two. Learning rates $\alpha$ and discount factors $\gamma$ follow the per-agent, per-noise settings selected in our prior work, reproduced in Table~\ref{tab:hyperparams} for self-containment. Exploration uses an $\varepsilon$-greedy schedule decayed from $1.0$ to $0.01$ over training, with exploration biased toward a $20\%$ attack rate so that unbiased $50\%$-rate exploration does not drive the QBER past threshold and cause universal early detection before learning occurs.

\begin{table}[t]
\centering
\caption{Attacker learning rate $\alpha$ and discount factor $\gamma$ by learning rule and channel noise $\mu_{ch}$, following our previous work \cite{bagunu2026adversarial}.}
\label{tab:hyperparams}
\renewcommand{\arraystretch}{1.2}
\begin{tabular}{llccc}
\toprule
 & & $\mu_{ch}=1\%$ & $\mu_{ch}=3\%$ & $\mu_{ch}=5\%$ \\
\midrule
\multirow{2}{*}{Q-Learning}
 & $\alpha$ & $0.50$ & $0.50$ & $0.50$ \\
 & $\gamma$ & $0.50$ & $0.50$ & $0.50$ \\
\midrule
\multirow{2}{*}{SARSA}
 & $\alpha$ & $0.01$ & $0.10$ & $0.50$ \\
 & $\gamma$ & $0.99$ & $0.99$ & $0.99$ \\
\midrule
\multirow{2}{*}{Double Q-Learning}
 & $\alpha$ & $0.50$ & $0.50$ & $0.50$ \\
 & $\gamma$ & $0.50$ & $0.50$ & $0.75$ \\
\bottomrule
\end{tabular}
\end{table}

\subsection{Distillable key fraction}
\label{sec:methods-key}
Detection rate alone does not capture the security consequence of an attack, so we also report the distillable secret-key fraction achievable at the eavesdropper's induced error rate. For an error rate $e$, the asymptotic secure fraction under one-way post-processing is
\begin{equation}
    r(e) = 1 - 2h(e),
    \label{eq:keyfrac}
\end{equation}
where $h(e)$ is the binary Shannon entropy \cite{shor2000simple},
\begin{equation}
    h(e) = -e\log_2 e - (1-e)\log_2(1-e).
    \label{eq:entropy}
\end{equation}
We clamp $r(e)$ to $[0,1]$, and it vanishes at $e = 11\%$ and above. For each evaluation episode we take $e$ to be the mean of that episode's completed-block QBERs. For a detected episode this includes the block on which the abort occurred, and we report the key fraction averaged over evaluation episodes. Because $r(e)$ is concave, averaging $r$ over episodes gives $\mathbb{E}[r(\bar{e})] \leq r(\mathbb{E}[\bar{e}])$, so our episode-averaged key fraction is a lower bound on the $r$ of the pooled QBER. The reported key fractions are therefore conservative toward the attacker rather than favorable to it. A successful stealthy attack is one that keeps the detection rate low and $r(e)$ high, so that the key remains usable while the eavesdropper gathers information. The two quantities are not interchangeable. A fixed-rate attacker operating just below the abort threshold induces an error rate near $e \approx 10.4\%$, at which the distillable key fraction is close to zero, whereas an attacker that holds the induced QBER near the background level retains most of it. Reporting both quantities prevents the misleading conclusion that a low detection rate alone implies a compromised key. The key fractions we report throughout are computed under greedy evaluation of the trained attacker, so they reflect the learned policy rather than its exploration behavior during training.

\subsection{The learning defender}
\label{sec:methods-def}
The defender is likewise a tabular RL agent, but it adapts the one lever available to a QBER-based detector which is the abort threshold. Its state encodes the announced block QBER ($16$ bins of width $1\%$) and the block index ($10$ bins), which gives $16 \times 10 = 160$ states. Its action selects the abort threshold from $\{0.08, 0.09, 0.10, 0.11, 0.12\}$, which aborts the current block if the block QBER meets or exceeds the chosen threshold. We deliberately restrict the defender to this single threshold lever since it is the minimal adaptive detector, and showing that even this minimal defender fails to settle the contest (Sec.~\ref{sec:results}) is a cleaner result than one obtained with a more elaborate detector. Richer levers are discussed as future work in Sec.~\ref{sec:conclusion}.

The defender is trained with a terminal, per-episode reward. On an attacked episode it receives $+10$ for aborting (correctly catching the attacker) and $-1$ for missing, while on an honest episode it receives $-10$ for aborting (a false alarm) and $+1$ for passing. The terminal reward is propagated to all threshold decisions taken within the episode.

\subsection{False-alarm discipline}
\label{sec:methods-fpr}
A detector that aborts more aggressively will catch more attackers only because it also raises more false alarms on honest traffic, so a defender trained purely to catch attacks would collapse to aborting almost everything. To prevent this, the defender is trained under a balanced mixture of honest and attacked episodes that charges it a direct cost for aborting honest traffic. Each defender training episode is honest with probability $\mathrm{honest\_frac}$ and attacked otherwise, and we use $\mathrm{honest\_frac}=0.5$ at every noise level, so that half of all training episodes carry no attack and a false abort on them is penalized. Because the false-alarm penalty is incurred only on honest episodes, the defender cannot abort freely, and it must separate attacked from honest traffic to gain reward. This imposes a uniform false-alarm discipline across all noise levels rather than a per-level calibration to a reference detector's false-positive rate. The discipline is identical across agents and rounds, so the comparisons in Sec.~\ref{sec:results} hold the defender's false-alarm pressure fixed while the learning dynamics vary.

\subsection{Alternating best-response game and exploitability}
\label{sec:methods-game}
We train the attacker and defender via alternating best response, in which one player is fully trained against a frozen opponent before the roles swap. We do not claim a Stackelberg solution, in which a leader optimizes while anticipating the follower's response. The alternation here simply freezes one player while the other best-responds, which is what encodes the last-mover structure we analyze. The game proceeds in rounds.
\begin{enumerate}
    \item \textbf{Round $0$.} An attacker is trained to best-respond to a static BB84 detector. This reproduces the attacker from our prior work.
    \item \textbf{Defender step.} The attacker is frozen and the learning defender is trained to best-respond to it, under the balanced honest and attacked training mixture.
    \item \textbf{Attacker step.} The defender is frozen and a fresh attacker is trained to best-respond to it.
    \item The defender and attacker steps repeat for $K$ rounds.
\end{enumerate}
Each training invocation uses $10{,}000$ episodes with a fixed horizon of $10$ blocks, and greedy evaluation uses $1{,}000$ episodes. Freezing one player while the other best-responds encodes the ``who moves last'' structure of the interaction and avoids the instability of two tabular learners updating simultaneously.

After $K$ rounds we measure the exploitability. We freeze the final defender and train a brand-new attacker against it, which reports that attacker's detection rate and distillable key fraction. Low detection together with high key fraction means a fresh attacker can still evade the final defender, so the window remains open, whereas high detection with low key fraction means the defender contains it. To diagnose convergence as opposed to non-convergence, we also record the fresh-attacker detection rate at every round and summarize its trajectory by its swing amplitude, taken as the standard deviation of the detection rate over the first versus the second half of the rounds. A trajectory that is settling has a shrinking second-half amplitude. A run is counted as settled when the standard deviation of its fresh-attacker detection rate over the final three rounds falls below $5$ percentage points. At $K=5$ this tail is only three points, so the settled count should be read as a coarse indicator of whether a trajectory has flattened rather than as a precise convergence test, and the $K=20$ analysis below is what establishes non-convergence.

A persistent amplitude establishes that the trajectory does not settle, but it does not by itself show that the round-to-round motion is a deterministic oscillation rather than a sequence of independent draws from a stationary distribution with nonzero variance. We therefore do not characterize the motion as a regular cycle. We report the lag-one autocorrelation of each trajectory, which is strongly negative for a true alternating oscillation and near zero for independent fluctuation, and we run a control that isolates how much of the swing is intrinsic to the best-response procedure. In the control we freeze the final defender of a cell and train five fresh attackers against it from different seeds, which then measures the spread of their detection rates. Retraining a best-responder against an unchanged opponent should give a small spread if the best-response procedure is near-deterministic, so a spread comparable to the round-to-round swing would show that the swing lives at the scale of best-response variance rather than reflecting a chase between evolving strategies. To test whether the defender nonetheless converges while the attacker alone fluctuates, we additionally record, at every round and for each state a greedy attacker reaches, the defender's abort-or-continue decision evaluated at that state's QBER. Then, we measure how often that decision changes between rounds. We restrict this to the states whose QBER lies in the adaptive band, where the chosen threshold can actually flip the decision, since outside that band every threshold gives the same outcome and a label change is inert. A defender that has settled would repeat the same decisions round to round. As a reference scale we also retrain the defender against a single frozen attacker across seeds and measure the spread of its decisions. This is the defender-side analogue of the attacker-variance control.

We run the full factorial of three learning rules, three noise levels $\mu_{ch} \in \{1\%, 3\%, 5\%\}$, and five random seeds at $K = 5$, and we extend a targeted set of cells to $K = 20$ with five seeds each to distinguish genuine non-convergence from slow convergence.

\section{Results}
\label{sec:results}

We present the results in four parts. We first establish the central finding, that the alternating best-response game does not settle (Sec.~\ref{sec:results-noconv}). We then characterize the noise-dependent behavior of the exploitability (Sec.~\ref{sec:results-phase}), relate detection to the security-relevant key fraction (Sec.~\ref{sec:results-tradeoff}), and finally confirm, with the extended $K=20$ runs and two best-response-variance controls, one per player, that the behavior is genuine non-convergence rather than slow convergence (Sec.~\ref{sec:results-converge}). All exploitability figures are computed against a fresh attacker best-responding to the final defender, and all comparisons hold the defender's false-alarm discipline fixed.

\subsection{The best-response game does not converge}
\label{sec:results-noconv}

\begin{figure*}[t]
\centering
\includegraphics[width=\textwidth]{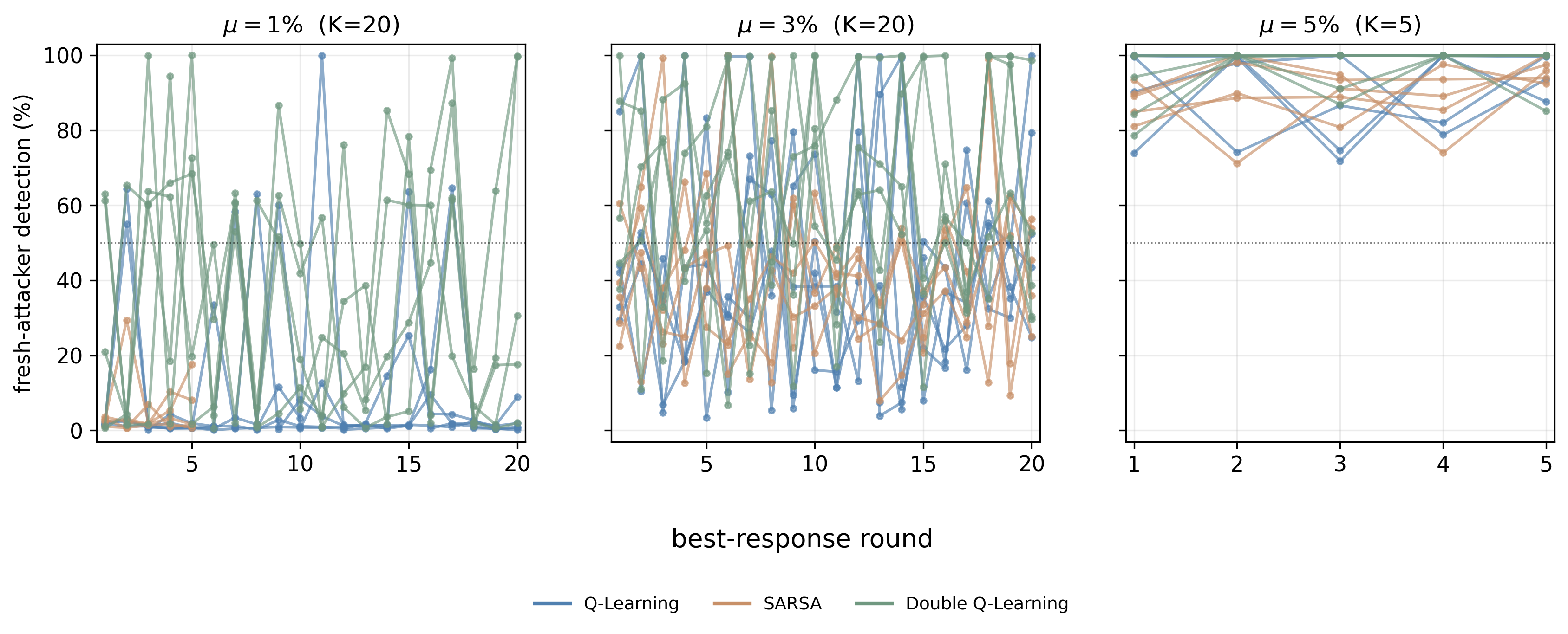}
\caption{Fresh-attacker detection rate across best-response rounds, for each learning rule (color) and channel noise level (panel). The contest does not settle. At $\mu_{ch}=3\%$ the detection rate swings across nearly the full range from round to round, at $\mu_{ch}=1\%$ the attacker is dominant but the detector mounts brief recoveries, and at $\mu_{ch}=5\%$ channel noise pins detection near $100\%$ independently of the learned policy. Each panel is annotated with the longest $K$ it shows. The $K=20$ trajectories are the targeted confirmation cells and the rest are the $K=5$ sweep.}
\label{fig:cycling}
\end{figure*}

Figure~\ref{fig:cycling} shows the fresh-attacker detection rate at every best-response round, for each learning rule and noise level. If the contest were settling, these trajectories would flatten as the rounds progress, as the detector and attacker approached a fixed point and the detection rate stopped moving. However, they do not. At $\mu_{ch}=3\%$ (center panel) every agent swings across nearly the full range of detection rates from one round to the next, round after round, with no sign of a stable value. The swings also include occasional rounds in which one side briefly finds a near-dominating policy before losing it again. At $\mu_{ch}=1\%$ (left panel) the trajectories spend most of their time near zero detection, with the attacker dominant, but are punctuated by sharp, short-lived surges in which a newly trained defender briefly catches the attacker before the next fresh attacker defeats it again. Only at $\mu_{ch}=5\%$ (right panel) are the trajectories flat, pinned near total detection, and there, as we discuss below, the channel rather than the learned policy is responsible.

The interaction therefore does not settle under best-response dynamics. The exploitability does not approach a fixed point but keeps fluctuating out to twenty rounds. This non-convergence is the central result of the paper, and the controls in Sec.~\ref{sec:results-converge} establish it from both sides. This shows that neither the attacker nor the defender converges to a fixed response. A reactive detector, however it adapts its threshold, cannot settle the contest against an adaptive attacker, and it can only trade places with it. We take up the mechanism we hold responsible, a last-mover information asymmetry, in Sec.~\ref{sec:discussion}, where the evidence for it is strongest at intermediate noise.

\subsection{A noise-dependent phase structure}
\label{sec:results-phase}

\begin{figure}[t]
\centering
\includegraphics[width=\columnwidth]{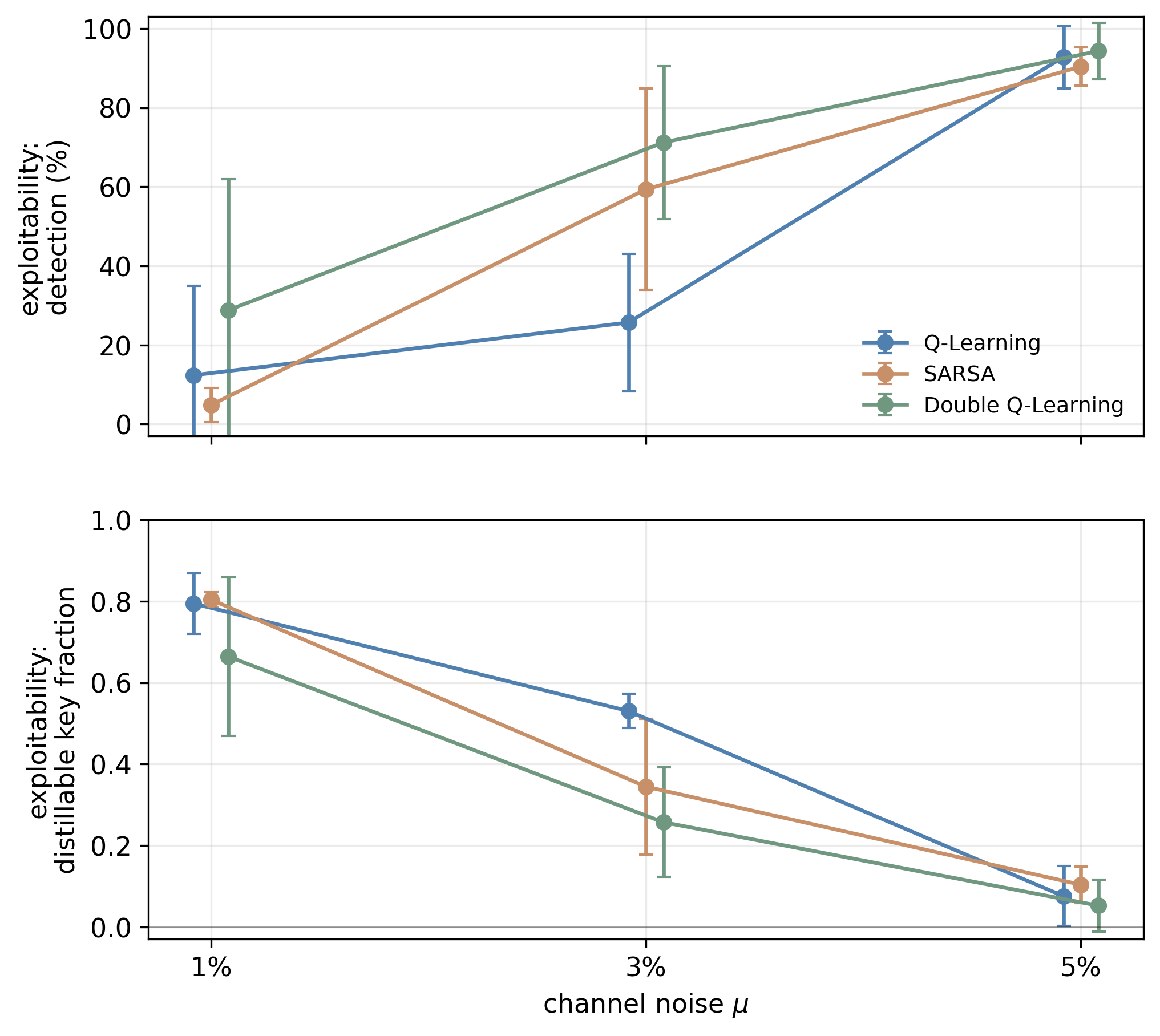}
\caption{Exploitability of the final defender versus channel noise, by learning rule (mean $\pm$ std over five seeds). Top is the fresh-attacker detection rate, while the bottom one is the distillable key fraction. The wide error bars at $\mu_{ch}=1\%$ reflect the non-convergence of Fig.~\ref{fig:cycling} rather than seed-to-seed measurement noise.}
\label{fig:gradient}
\end{figure}

Table~\ref{tab:exploit} reports the exploitability of the final defender for every agent and noise level, and Fig.~\ref{fig:gradient} shows the same data as a function of noise. Two things vary with the channel, namely whether a fresh attacker can still win on average, and whether the game settles at all. These do not coincide, and separating them is essential to reading the result correctly.

Whether a fresh attacker wins on average is strongly noise-dependent. At $\mu_{ch}=1\%$ the window is open. When averaged over agents, a fresh attacker retains a key fraction of $0.754$ at only $15.3\%$ detection, so the learned defender does not close the stealth window at the low noise where practical QKD operates. At $\mu_{ch}=3\%$ the average shifts to the defender ($52.1\%$ detection, $0.377$ key fraction), and at $\mu_{ch}=5\%$ the attacker is contained entirely ($92.5\%$ detection, $0.077$ key fraction). The containment at $\mu_{ch}=5\%$, however, is not a property of the learned policy. The cleanest evidence for this is a direct comparison with the static detector of our prior work. Against the fixed $11\%$ abort threshold, a learned attacker at $\mu_{ch}=5\%$ is already detected at $91.7\%$, $93.6\%$, and $97.3\%$ for Q-Learning, SARSA, and Double Q-Learning respectively \cite{bagunu2026adversarial}. The learned defender here reaches $92.8\%$, $90.4\%$, and $94.3\%$ on the same three agents, which is statistically indistinguishable and in two of three cases slightly lower. The learned defender therefore adds essentially nothing at $\mu_{ch}=5\%$. The attack is already suppressed by channel statistics before the defender learns anything, so we do not read $\mu_{ch}=5\%$ as a defender victory.

Whether the game settles tells a different and more uniform story. The fraction of seeds that reach a settled trajectory is high only at $\mu_{ch}=1\%$, and even there it is agent-dependent. Q-Learning settles in all five seeds and SARSA in four, but Double Q-Learning settles in only one. At $\mu_{ch}=3\%$ no run of any agent settles. This is the critical point for interpreting the averages above, because the defender's apparent win at $\mu_{ch}=3\%$ is not a stable outcome but one draw from a distribution that the dynamics never stop sampling, as the next subsection confirms. At a fixed noise level, then, the attacker's learning rule governs the dynamics. Double Q-Learning's bias-corrected value estimates make it a sharper best-responder that denies the defender a stable foothold, whereas Q-Learning is more readily pinned to a settled attacker-dominant state.

\begin{table}[t]
\centering
\caption{Exploitability of the final learned defender, giving fresh-attacker detection rate and distillable key fraction, mean $\pm$ std over five seeds. ``Settled'' counts seeds whose best-response trajectory reached a flat tail.}
\label{tab:exploit}
\renewcommand{\arraystretch}{1.2}
\begin{tabular}{llccc}
\toprule
Agent & $\mu_{ch}$ & Detection (\%) & Key fraction & Settled \\
\midrule
\multirow{3}{*}{Q-Learning}
 & $1\%$ & $12.4 \pm 22.6$ & $0.794 \pm 0.074$ & $5/5$ \\
 & $3\%$ & $25.7 \pm 17.4$ & $0.530 \pm 0.042$ & $0/5$ \\
 & $5\%$ & $92.8 \pm\phantom{0}7.9$ & $0.076 \pm 0.074$ & $0/5$ \\
\midrule
\multirow{3}{*}{SARSA}
 & $1\%$ & $\phantom{0}4.8 \pm\phantom{0}4.4$ & $0.804 \pm 0.018$ & $4/5$ \\
 & $3\%$ & $59.4 \pm 25.5$ & $0.344 \pm 0.167$ & $0/5$ \\
 & $5\%$ & $90.4 \pm\phantom{0}4.8$ & $0.104 \pm 0.045$ & $2/5$ \\
\midrule
\multirow{3}{*}{Double Q-Learning}
 & $1\%$ & $28.8 \pm 33.1$ & $0.664 \pm 0.194$ & $1/5$ \\
 & $3\%$ & $71.2 \pm 19.3$ & $0.257 \pm 0.135$ & $0/5$ \\
 & $5\%$ & $94.3 \pm\phantom{0}7.1$ & $0.052 \pm 0.064$ & $3/5$ \\
\bottomrule
\end{tabular}
\end{table}

\subsection{Detection versus usable key}
\label{sec:results-tradeoff}

\begin{figure}[t]
\centering
\includegraphics[width=\columnwidth]{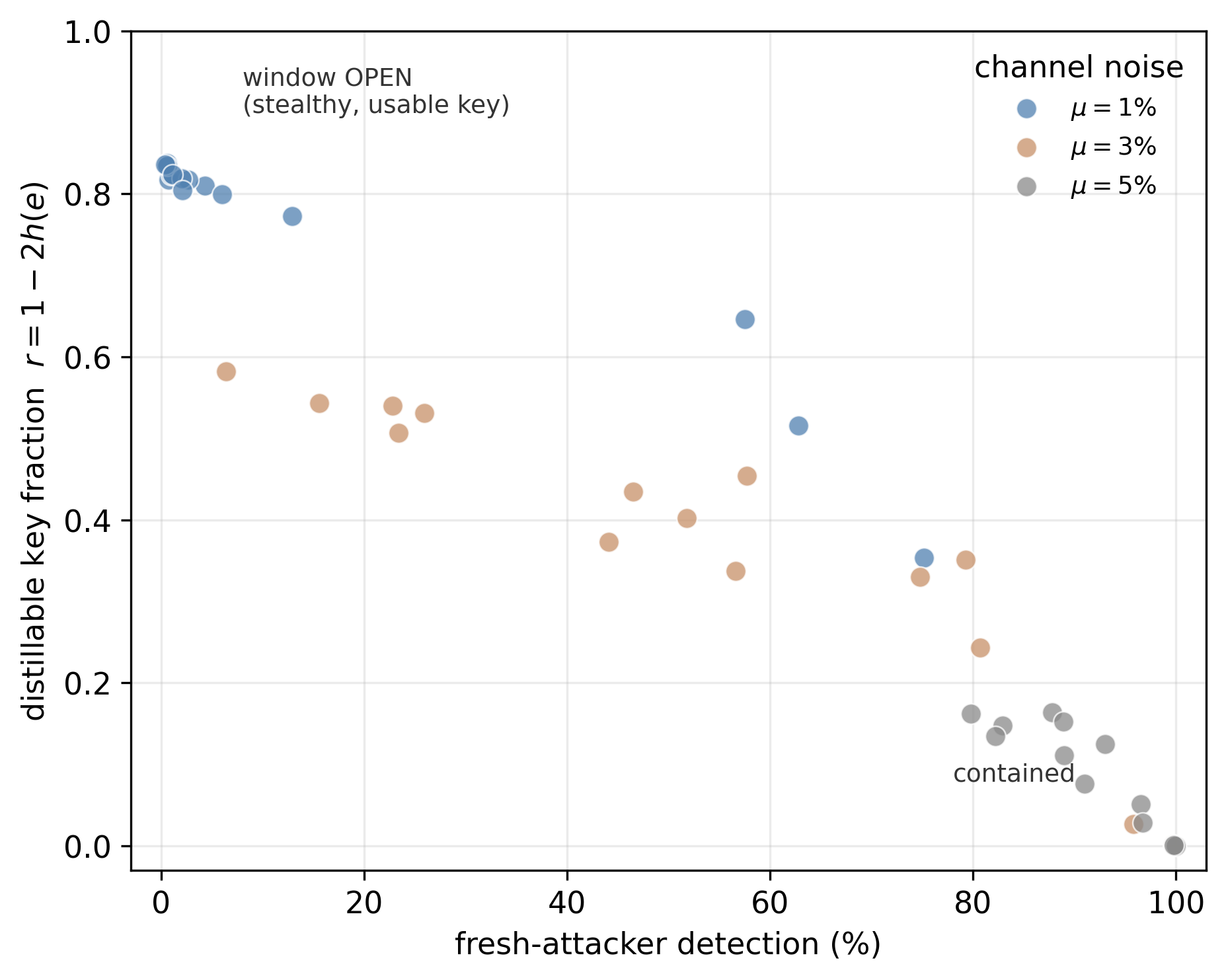}
\caption{Distillable key fraction $r(e)=1-2h(e)$ versus fresh-attacker detection rate, one point per run, colored by channel noise. Upper-left is an open window (stealthy, usable key), and lower-right is containment. The regimes separate cleanly, and the $\mu_{ch}=1\%$ spread along the horizontal axis reflects the non-converging dynamics.}
\label{fig:tradeoff}
\end{figure}

Detection rate and key fraction are related but not interchangeable. The security-relevant quantity is the key fraction, since a low detection rate matters only if the key the attacker obtains is distillable. Figure~\ref{fig:tradeoff} plots the two against each other for every run. The three noise regimes occupy distinct regions of the plane. The $\mu_{ch}=1\%$ runs cluster in the upper-left (low detection, high key fraction, the window open), the $\mu_{ch}=5\%$ runs collapse to the lower-right (high detection, no usable key, contained by the channel), and the $\mu_{ch}=3\%$ runs scatter along the diagonal between them. The spread of the $\mu_{ch}=1\%$ cluster along the detection axis is, once again, the non-convergence made visible, as individual runs land at different points of the fluctuating dynamics.

\subsection{Non-convergence, confirmed at \texorpdfstring{$K=20$}{K=20}}
\label{sec:results-converge}

The $K=5$ runs show large swings, but five rounds are too few to distinguish non-convergence from a slow convergence that has not yet stabilized. To settle this we extended a targeted set of cells to $K=20$ rounds with five seeds each, which is enough to compare the first ten rounds against the last ten. For each cell we measure the swing amplitude, the standard deviation of fresh-attacker detection, over the first half of the rounds and over the second. If the game were slowly converging, the second-half amplitude would be markedly smaller than the first. Instead, it is unchanged, which is the characteristic of non-convergence rather than slow convergence.

\begin{figure}[t]
\centering
\includegraphics[width=\columnwidth]{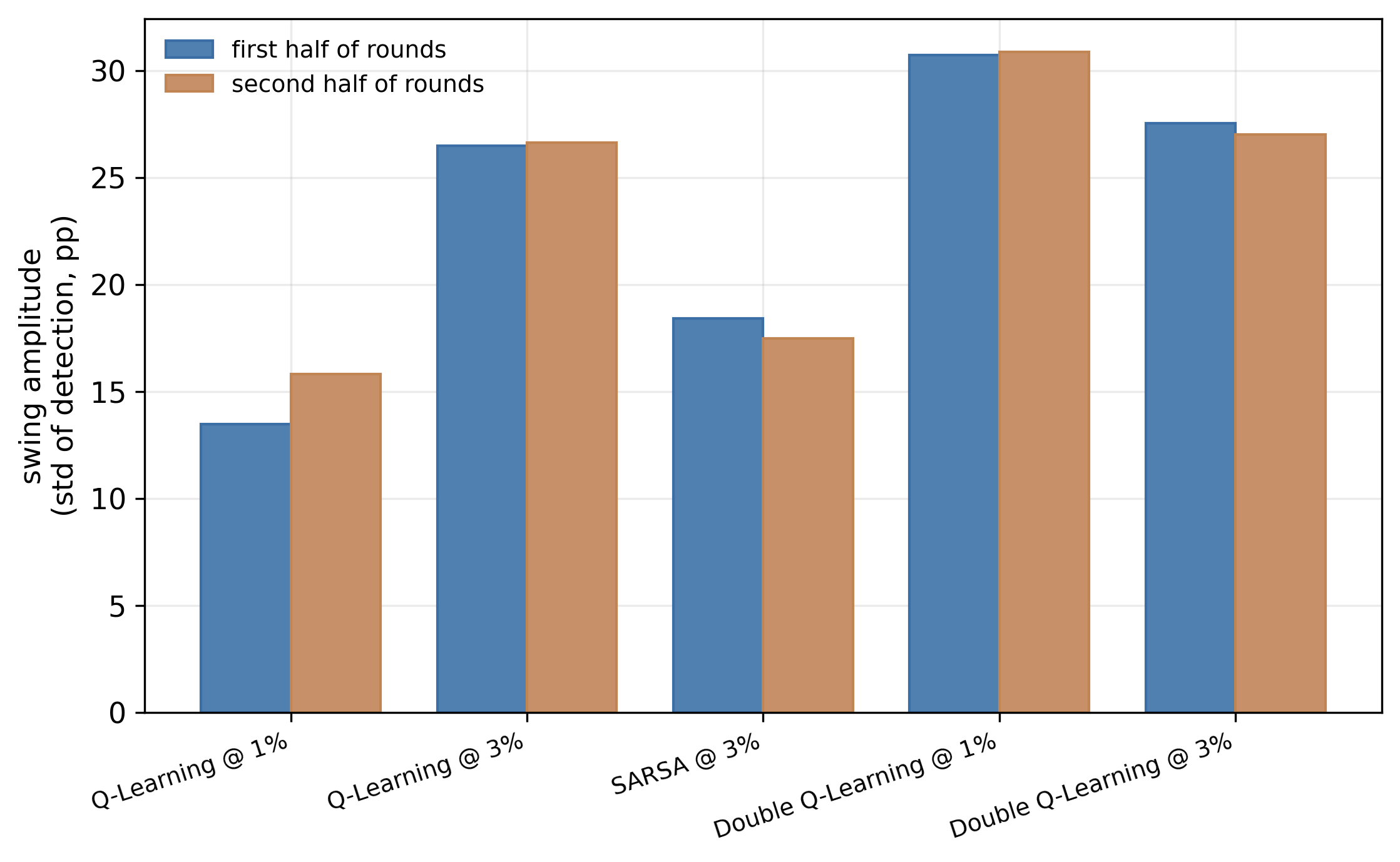}
\caption{Swing amplitude over the first versus the second half of $K=20$ rounds, per confirmation cell (five seeds). In every cell the second-half amplitude is comparable to the first and in none does it decay toward zero, which confirms genuine non-convergence rather than slow convergence.}
\label{fig:amplitude}
\end{figure}

Table~\ref{tab:converge} reports the result. In every cell the second-half amplitude remains large, between $16$ and $31$ percentage points, and in no cell does it decay toward zero. At $\mu_{ch}=3\%$ it is essentially unchanged from the first half, which holds near $24$ percentage points averaged over the three learning rules. Even the two $\mu_{ch}=1\%$ cells, where some seeds settle, retain a second-half amplitude comparable to the first once all five seeds are included. The amplitude is comparable across the two halves in every cell and decays in none, which is the direct signature of non-convergence rather than slow convergence. Figure~\ref{fig:amplitude} shows the first- against second-half amplitudes directly, and in every cell the second-half bar is comparable in height to the first.

The persistent amplitude shows the trajectory does not settle, but it does not by itself reveal whether the round-to-round motion carries any structure or is a sequence of independent draws. The lag-one autocorrelation $\rho_1$ separates the two. A deterministic alternating oscillation drives $\rho_1$ strongly negative, independent draws leave it near its small-sample null, and a weak reactive effect sits in between as mild anti-persistence. For an independent sequence of length $20$ the null value of $\rho_1$ is approximately $-1/20 = -0.05$, with a per-seed standard error near $0.22$, so the five-seed mean has a standard error near $0.10$.

Measured against this null, the behavior splits sharply by noise. At $\mu_{ch}=3\%$ the mean $\rho_1$ is $-0.27$ for Q-Learning, $-0.33$ for SARSA, and $-0.32$ for Double Q-Learning. Each are roughly $2.3$ to $2.8$ standard errors below the null, which is significant anti-persistence. The round after a defender-favorable round tends to swing back toward the attacker, which is the characteristic of a reactive advantage that changes hands. At $\mu_{ch}=1\%$, by contrast, the mean $\rho_1$ is $-0.02$ for Q-Learning and $-0.16$ for Double Q-Learning, which are within one standard error of the null and therefore indistinguishable from independent best-response noise. The reversal rates tell the same story, which lies between $0.58$ and $0.78$, inside the $[0.44, 0.83]$ band expected for an independent sequence of this length. We therefore describe the behavior as non-convergence with round-to-round fluctuation, structured by a significant anti-persistence at intermediate noise and indistinguishable from best-response noise at low noise, rather than as a regular limit cycle.

A control experiment pins down the scale of that fluctuation. For each confirmation cell we froze the final defender and trained five fresh attackers against it from different seeds, so that the only source of variation is the best-response procedure with no evolving opponent. Table~\ref{tab:control} reports the resulting spread. Retraining the attacker against an unchanged defender already produces a standard deviation of between $8$ and $39$ percentage points in detection, which is comparable to, and in several cells larger than, the round-to-round swing amplitudes of Table~\ref{tab:converge}. The round-to-round motion is therefore of the same magnitude as the intrinsic variance of best-responding, so the game fluctuates at the scale of its own best-response variance rather than chasing a moving target through a regular orbit. This is the precise sense in which the contest does not settle. It does not approach any fixed point, and the variation that remains is at the noise floor of the best-response procedure itself. The exploitability standard deviations in Table~\ref{tab:exploit}, which come from a single re-seeded attacker against each final defender, are one draw from exactly this distribution and fall inside the ranges the control reports.

The control shows that the attacker side keeps moving, but it leaves open whether the defender has quietly settled. The exploitability swing then reflects only attacker retraining variance around a fixed defender. We examine the defender directly, and we are careful to measure what the defender does rather than the label it carries. For each confirmation cell we reconstructed the defender at every round and recorded, for each state a greedy attacker reaches, the resulting abort-or-continue decision evaluated at that state's QBER rather than the chosen threshold index. The decision is what matters. For a QBER bin below the lowest threshold every threshold continues, and for a bin above the highest every threshold aborts, so a threshold change in those states is inert. Only states whose QBER falls in the adaptive band carry a decision the threshold can flip, and we measure how often the defender's decision on those states changes from one round to the next. Between $18\%$ and $39\%$ of these decisions change per round (Table~\ref{tab:defender}), so the defender's behavior is not static. However, the label-level figure of roughly $75\%$ to $80\%$ overstates this because it counts the inert changes.

To calibrate that round-to-round change we run a defender-side control analogous to the attacker control, where we retrain the defender against a single frozen attacker across five seeds and measure the pairwise disagreement of its decisions. Table~\ref{tab:defender} reports both. The round-to-round decision change and the frozen-attacker re-seed spread are of the same magnitude in every cell, and in three of the five the round-to-round change is the smaller of the two. The exception in direction is Double Q-Learning at $\mu_{ch}=1\%$, where the re-seed spread ($60\%$) exceeds the round-to-round change ($39\%$). This is the cell whose defender training is least reproducible across seeds, which widens the re-seed spread rather than signaling any extra round-to-round stability. In both directions the conclusion is the same, since the defender's decisions move about as much when the opponent is held fixed as they do across the game. So the defender, like the attacker, fluctuates at the scale of its own best-response variance rather than tracking an evolving opponent to a settled response. Both controls point to the same conclusion that the exploitability does not approach a fixed point, and the residual motion on each side is at the noise floor of the best-response procedure. This is the precise sense in which the contest does not settle, and it is why we do not claim a stable equilibrium of the strategy profile.

\begin{table}[t]
\centering
\caption{Swing amplitude (std of fresh-attacker detection, in percentage points) over the first versus second half of $K=20$ rounds, the round-to-round reversal rate, and the mean lag-one autocorrelation $\rho_1$, for the targeted confirmation cells (mean over five seeds). Reversal rates lie within the $[0.44,0.83]$ band expected for an independent sequence. $\rho_1$ is significantly negative at $\mu_{ch}=3\%$, which indicates anti-persistence, and indistinguishable from the null at $\mu_{ch}=1\%$.}
\label{tab:converge}
\renewcommand{\arraystretch}{1.2}
\begin{tabular}{llcccc}
\toprule
Agent & $\mu_{ch}$ & 1st-half & 2nd-half & Reversal & $\rho_1$ \\
\midrule
Q-Learning        & $1\%$ & $13.4$ & $15.8$ & $0.58$ & $-0.02$ \\
Q-Learning        & $3\%$ & $26.5$ & $26.6$ & $0.78$ & $-0.27$ \\
SARSA             & $3\%$ & $18.4$ & $17.5$ & $0.75$ & $-0.33$ \\
Double Q-Learning & $1\%$ & $30.7$ & $30.9$ & $0.66$ & $-0.16$ \\
Double Q-Learning & $3\%$ & $27.5$ & $27.0$ & $0.72$ & $-0.32$ \\
\bottomrule
\end{tabular}
\end{table}

\begin{table}[t]
\centering
\caption{Best-response-variance control. For each confirmation cell the final defender is frozen and five fresh attackers are trained against it from different seeds. The resulting spread (mean $\pm$ std of detection, in percentage points) is comparable to the round-to-round swing amplitudes of Table~\ref{tab:converge}, so the game fluctuates at the scale of its own best-response variance.}
\label{tab:control}
\renewcommand{\arraystretch}{1.2}
\begin{tabular}{llc}
\toprule
Agent & $\mu_{ch}$ & Detection (\%) \\
\midrule
Q-Learning        & $1\%$ & $22.9 \pm 38.6$ \\
Q-Learning        & $3\%$ & $79.7 \pm 25.4$ \\
SARSA             & $3\%$ & $32.9 \pm\phantom{0}8.3$ \\
Double Q-Learning & $1\%$ & $22.9 \pm 23.6$ \\
Double Q-Learning & $3\%$ & $57.2 \pm 35.5$ \\
\bottomrule
\end{tabular}
\end{table}

\begin{table}[t]
\centering
\caption{Defender decision dynamics. ``Round-to-round'' is the fraction of decision-relevant reachable states (QBER in the adaptive band) whose abort-or-continue decision changes between successive best-response rounds. ``Re-seed spread'' is the pairwise disagreement of the same decisions when the defender is retrained against a single frozen attacker across five seeds. The two are of the same magnitude in every cell, so the defender's round-to-round motion is at the scale of its own best-response variance.}
\label{tab:defender}
\renewcommand{\arraystretch}{1.2}
\begin{tabular}{llcc}
\toprule
Agent & $\mu_{ch}$ & Round-to-round (\%) & Re-seed spread (\%) \\
\midrule
Q-Learning        & $1\%$ & $38.6$ & $33.3$ \\
Q-Learning        & $3\%$ & $18.1$ & $20.0$ \\
SARSA             & $3\%$ & $23.5$ & $18.9$ \\
Double Q-Learning & $1\%$ & $39.0$ & $60.0$ \\
Double Q-Learning & $3\%$ & $17.5$ & $22.4$ \\
\bottomrule
\end{tabular}
\end{table}

Taken together, the results establish three things. First, the learned eavesdropping and detection game does not settle under best-response dynamics out to twenty rounds, with the exploitability fluctuating at the scale of best-response variance and neither player converging to a fixed response. Second, whether a fresh attacker wins on average is governed by channel noise, and the window stays open at the low noise where practical QKD operates. Third, the defender’s apparent advantage at intermediate noise is not a stable outcome but one draw from a distribution that the dynamics keep resampling.

\section{Discussion}
\label{sec:discussion}

We now interpret the non-convergence result. We consider what it means for the defensibility of BB84 against an adaptive attacker (Sec.~\ref{sec:disc-mean}), the structural cause we hold responsible (Sec.~\ref{sec:disc-mech}), why we expect it to survive a more capable defender (Sec.~\ref{sec:disc-robust}), the limitations that bound the claim (Sec.~\ref{sec:disc-limits}), and what the result implies for how such systems should be defended (Sec.~\ref{sec:disc-defense}).

\subsection{What non-convergence means for BB84 detection}
\label{sec:disc-mean}

The question that motivated this work was whether a learning detector can close the stealth window that a learning attacker opens. Our previous work \cite{bagunu2026adversarial} showed that a fixed detector cannot. A natural expectation is that an equally adaptive detector should be able to, by tracking the attacker and tightening its response. The results show that this expectation is misplaced, but not in the way one might assume. It is not that the defender always loses, since at intermediate and high noise a fresh attacker is in fact contained on average. It is that the contest does not settle. The defender can best-respond to the current attacker, but a fresh attacker then best-responds to that defender, and the advantage changes hands. Out to twenty rounds of adaptation the swing in exploitability does not decay, which remains as large in the second half of the run as in the first (Table~\ref{tab:converge}). Its magnitude is set by the best-response variance rather than by any convergence toward a fixed point. The two controls make this concrete, and they show that the attacker's and the defender's residual motion is each at the scale of their own best-response variance (Tables~\ref{tab:control} and~\ref{tab:defender}). Therefore, neither side settles to a fixed response.

The practical reading is a caution against a particular kind of security argument. One cannot certify an adaptive detector by training it against the current best attacker and reporting that it contains that attacker, because the next attacker to come along is trained against the detector, not the attacker it replaced. A snapshot of the contest at any single round, including a favorable one, is not evidence that the detector has closed the window. The security-relevant summary is the dynamics, not the snapshot, and the dynamics do not settle. At the low noise where deployed QKD operates ($\mu_{ch}=1\%$), the snapshot is in any case unfavorable, since a fresh attacker retains roughly three-quarters of the distillable key (Table~\ref{tab:exploit}).

\subsection{The last-mover information asymmetry}
\label{sec:disc-mech}

As the candidate explanation for the non-convergence we examine a structural asymmetry in the information available to the two players, which the trajectory statistics support at intermediate noise and leave open at low noise. The block QBER is broadcast publicly over the authenticated classical channel. It is the only signal either player has about the other's current behavior, and both condition on it. In the alternating best response, whichever player trains against the most recently revealed QBER is responding to the other's committed strategy, while the other has not yet seen the response. That player, the last mover, holds a temporary advantage for the current round. Because the roles of first and last mover swap at every adaptation, the advantage swaps with them. This picture is clearest at intermediate noise, where the exploitability is significantly anti-persistent. At low noise the swapping is masked by the floor effect and the fluctuation is at the scale of best-response noise.

\begin{figure*}[t]
\centering
\includegraphics[width=\textwidth]{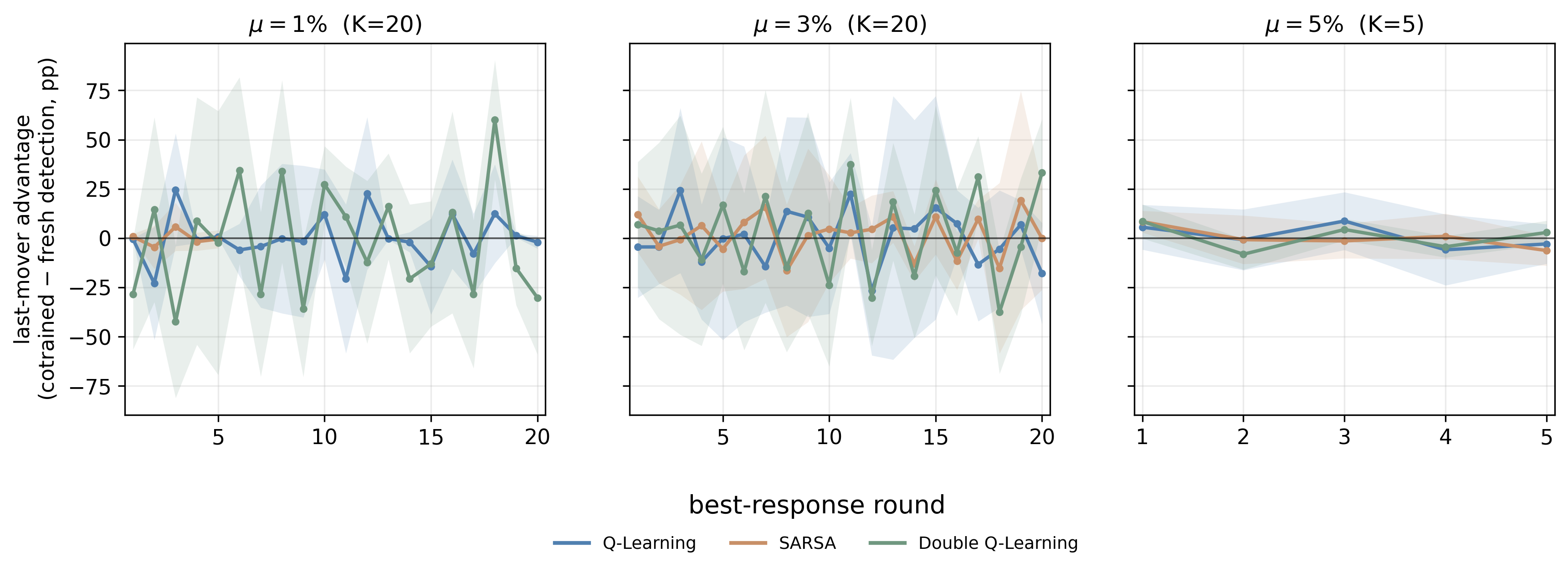}
\caption{The last-mover advantage per best-response round, defined as the difference between the frozen attacker’s detection against the newly trained defender (defender moved last) and a fresh attacker’s detection against it (attacker moved last). A positive value means the last mover wins that round. Lines are the seed mean and bands span $\pm$ one standard deviation across seeds, with color denoting the learning rule and the panel the channel noise. The signal is noisy rather than a constant gap, positive on average at $\mu_{ch}=3\%$ but with a band that crosses zero, and near zero at $\mu_{ch}=1\%$ and $5\%$. The stronger evidence for the last-mover reading is the anti-persistence of the exploitability trajectory at $\mu_{ch}=3\%$ (Table~\ref{tab:converge}), not this direct gap.}
\label{fig:lastmover}
\end{figure*}

The best-response logs let us probe this behavior directly. At each round we record two detection rates against the newly trained defender, that of the frozen attacker it was trained to catch (the defender moved last) and that of a fresh attacker trained afterward against it (the attacker moved last). Their difference is the last mover’s advantage for that round. Figure~\ref{fig:lastmover} plots this difference across rounds. The picture is noisy and noise-dependent rather than a clean constant gap. At $\mu_{ch}=3\%$ the difference is positive on average, which is consistent with the last mover holding an edge. However, it fluctuates widely from round to round and its seed band crosses zero, so the per-round advantage is a tendency rather than a guarantee. At $\mu_{ch}=1\%$ the difference hovers around zero which is expected under a floor effect, since both the stale and the fresh attacker are usually near zero detection at this noise and their difference is therefore small regardless of who moved last. At $\mu_{ch}=5\%$ it is also near zero there because the channel rather than the move order decides the outcome. The direct per-round gap is thus weak evidence on its own. The stronger evidence for the last-mover reading is the significant anti-persistence of the exploitability trajectory at $\mu_{ch}=3\%$ reported in Sec.~\ref{sec:results-converge}, which is what a reactive advantage that changes hands would produce, and which is absent at $\mu_{ch}=1\%$.

This framing explains the noise dependence without additional assumptions. At low noise the attacker's safe operating margin below the $11\%$ threshold is wide, so when the attacker moves last it can exploit the margin almost fully, and keep the attacker-favorable rounds close to full stealth. As noise rises the margin narrows, the attacker’s last-mover gain shrinks, and the center of mass of the fluctuation shifts toward the defender. By $\mu_{ch}=5\%$ the margin is so small that channel statistics alone push most blocks over threshold, regardless of either policy. The asymmetry is the same throughout, and what changes with noise is only how much the last mover can extract from it.

This picture is consistent with the attacker’s learning rule, not only the noise, shaping the dynamics. A sharper best-responder exploits its last-mover turn more completely and leaves the opponent less to recover on the next turn, which sustains the fluctuation rather than dampen it. This is why Double Q-Learning, whose bias-corrected value estimates make it the most accurate of the three best-responders, is the agent least likely to let the game settle, even at $\mu_{ch}=1\%$ where Q-Learning is frequently pinned to a stable attacker-dominant state (Table~\ref{tab:exploit}).

\subsection{Why a more capable defender is unlikely to help}
\label{sec:disc-robust}

Our defender adapts a single lever, which is the abort threshold. A reasonable objection is that a richer defender, one that also randomizes block sizes or checkpoint timing, or conditions on finer features of the error pattern, might close the window where the threshold alone cannot. We argue that it is unlikely to, and the reason is worth stating carefully, because the instinct that non-convergence signals an underpowered model is misleading here.

The behavior we observe is not the failure mode of an underpowered defender. An underpowered defender would lose consistently, with the game converging to an attacker-dominant fixed point. On the other hand an overpowered one would win consistently, converging to a defender-dominant fixed point. Neither is what the data shows. The exploitability instead fails to settle and the swing amplitude does not decay (Table~\ref{tab:converge}), and at intermediate noise it carries the significant anti-persistence that a reactive feedback loop would produce. The defender does not settle to a fixed response either, though its decisions move at the scale of its own best-response variance rather than more (Table~\ref{tab:defender}). Both players remain in motion rather than one having found a fixed response the other merely orbits. Non-convergence with this structure points to the interaction rather than to insufficient capacity, and adding capacity to a system that fails to settle generally reshapes the fluctuation rather than removing it. We note that this reading is best supported at $\mu_{ch}=3\%$, where the anti-persistence is significant, and is weaker at $\mu_{ch}=1\%$, where the fluctuation is indistinguishable from best-response noise.

The block QBER is broadcast publicly, both players condition on it, and whichever player adapts to it last holds the advantage for that round. A more capable defender is still a reactive defender, so it still responds to the broadcast QBER, and a fresh attacker trained afterward still sees the consequences of that response and adapts to them. The defender's sophistication becomes one more thing the next attacker best-responds to. Adding levers changes what each phase of the interaction looks like, but it does not change who moves last. If the move order rather than the move set is what keeps the contest from settling, as the intermediate-noise anti-persistence suggests, then enlarging the move set would not help. This is consistent with the classical theory, in which best-response dynamics can fail to converge regardless of the richness of the strategy space \cite{shapley1964some, mertikopoulos2018cycles}. Therefore, enlarging the defender's strategy space has no general reason to force convergence. We note also that the failure of reactivity is not peculiar to the learning defender, since in our previous work the static $11\%$ abort threshold also fails to close the window against the learned attacker. This is what motivated giving the detector the ability to learn in the first place.

We emphasize what this argument does and does not establish. We demonstrate non-convergence directly only for the minimal threshold defender, and the claim that richer reactive defenders also fail to settle is an argument from the last-mover asymmetry rather than an exhaustive experiment. Testing it directly across a family of higher-capacity reactive defenders is a natural next step, which we return to in Sec.~\ref{sec:conclusion}. What breaks the loop, by contrast, is not a better reactor but a defender that changes the information structure itself, and we take up that constructive direction in Sec.~\ref{sec:disc-defense}.

\subsection{Limitations and scope}
\label{sec:disc-limits}

Several boundaries bound the claim. First, and most importantly, our evidence for non-convergence is obtained under alternating best response, in which one player is fully trained against a frozen opponent before the roles swap. This is a standard and well-motivated training scheme for coupled learners \cite{brown1951iterative, busoniu2008comprehensive}, and it directly encodes the last-mover structure we analyze, but it is one dynamics among several. Whether the non-convergence is intrinsic to the game, persisting under simultaneous updates, fictitious play, or other learning dynamics, or is in part an artifact of the alternating protocol, is a question our experiments do not settle. We return to it as the primary direction for future work in Sec.~\ref{sec:conclusion}.

Second, the agents and environment are deliberately simple, with tabular value-based learners on a discrete state space, an idealized intercept-resend channel, and a single adaptive lever per side. As in our prior work, this transparency is a choice, since every learned value is inspectable, so the non-convergence can be attributed to the game's structure rather than to the opacity of a function approximator. However, it leaves open whether richer learners on a more realistic channel would show the same behavior. The defender in particular is trained with a coarse credit assignment, in which a single terminal episode reward is applied to every threshold decision taken within the episode. A pass early in an attacked episode that is caught only later still receives the catch reward. This scheme is transparent and well suited to the short horizon, but we cannot rule out that it contributes some of the instability we otherwise attribute to the last-mover asymmetry. A defender with finer per-decision credit assignment is a natural control for future work. Relatedly, the best-response-variance control of Table~\ref{tab:control} shows that a substantial part of the round-to-round variation is intrinsic to the best-response procedure itself. We read this not as a confound but as the correct scale at which to state the result, since it is precisely why we describe the dynamics as fluctuating at best-response-variance scale rather than executing a regular cycle. 

Third, our noise grid is coarse ($\mu_{ch}\in\{1\%,3\%,5\%\}$) and our seed count modest. This is adequate to establish the qualitative non-convergence and its noise dependence, but not to resolve fine structure near the transitions between regimes. We have therefore not sought to attach significance tests to the between-cell differences.

\subsection{Implications for defense}
\label{sec:disc-defense}

If reactivity cannot settle the contest, the implication is that defending against an adaptive eavesdropper is not a matter of out-adapting it. A detector that only reacts to the broadcast QBER is always a move behind, and making it a faster or more elaborate reactor does not change the order of information. The constructive direction suggested by the last-mover analysis is to break the asymmetry rather than chase it, by denying the attacker a clean, predictable signal to condition its last move on. Randomizing the checkpoint structure is one concrete step in this direction, though our previous work found that randomized checkpoints remove a specific temporal exploit without closing the underlying stealth window. Therefore, it is unlikely to be sufficient on its own. We do not evaluate such defenses here, since the contribution of this paper is the negative result that motivates them, namely that reactive detection, however adaptive, trades places with the attacker rather than defeating it. More broadly, the result argues for assessing adaptive-adversary defenses by their learning dynamics, by asking whether the contest converges and to whose advantage, rather than by their performance against a single fixed opponent.

\section{Conclusion}
\label{sec:conclusion}

We asked whether a learning detector can close the stealth window that a learning eavesdropper opens in BB84, and we answered it by casting the interaction as an alternating best-response game and measuring exploitability across rounds. The answer is that the question has no stable answer, since out to twenty best-response rounds the contest does not settle. The swing in a fresh attacker's exploitability against the final defender does not decay, which remains as large late in the run as early in it. Additionally, two controls, one retraining the attacker against a single frozen defender and one retraining the defender against a single frozen attacker, show this swing is of the same magnitude as the best-response procedure's own variance on each side. The contest fluctuates at that scale rather than approaching a fixed point, and neither player converges to a fixed response. A reactive detector can best-respond to the current attacker, but the next attacker best-responds to the detector, and the contest does not resolve. At intermediate noise this takes the form of an advantage that changes hands, which is reflected in the significant anti-persistence of the exploitability. At low noise the residual motion is indistinguishable from best-response noise.

The behavior is governed by channel noise through a single mechanism. At the low noise where practical QKD operates, the window is open, and a fresh attacker retains roughly three-quarters of the distillable key against the final learned defender. At intermediate noise the defender contains the attacker on average, but never stably, as one draw from a distribution the dynamics keep resampling rather than as a fixed outcome. At high noise the attacker is suppressed by the channel rather than by the learned policy. As the mechanism we examine a last-mover information asymmetry, in which, because the block QBER is broadcast publicly, whichever player adapts to it last holds a temporary advantage that alternates with the order of adaptation. The trajectory statistics support this at intermediate noise, where the exploitability is significantly anti-persistent. They leave it open at low noise, where the fluctuation is indistinguishable from the variance of the best-response procedure itself.

The implication is that defending against an adaptive eavesdropper is not a matter of out-adapting it. Any detector that only reacts to the broadcast QBER inherits the same asymmetry, so a more capable reactive detector should be expected to fail to settle as well, and closing the window calls for breaking the information asymmetry rather than chasing it. Three directions follow from this. The first, and most important, is to determine whether the non-convergence is intrinsic to the game or in part an artifact of the alternating best-response dynamics under which we observe it. Reproducing it under other learning dynamics, and connecting it to the known non-convergence of best response in classes of games \cite{shapley1964some, mertikopoulos2018cycles}, would settle this. The second is to test the finding against richer models on both sides, with higher-capacity defenders that have additional levers and attackers using function approximation on channels that include loss and detector imperfection. Our analysis predicts that any purely reactive defender will still fail to settle. The third is the constructive counterpart to our negative result, namely defenses that break the last-mover asymmetry rather than react within it, and whether any such defense closes the window that reactive detection cannot. The present contribution is to show, with directly inspectable classical learners, that reactive detection does not settle the adaptive eavesdropping contest, and that it only trades places within it.


\end{document}